\documentclass[12pt]{article}
\begin{document}
\title{Radial Oscillations of Neutron Stars in Strong Magnetic Fields}
\author{V.K.Gupta \footnote{E--mail : vkg@ducos.ernet.in} ,Vinita Tuli, S.Singh\footnote{E--mail : santokh@ducos.ernet.in}, J.D.Anand\footnote{E--mail : jda@ducos.ernet.in} and Ashok Goyal\footnote{E--mail : agoyal@ducos.ernet.in} \\
            {\em Department of Physics and Astrophysics,} \\
            {\em University of Delhi, Delhi-110 007 , India.} \\
            {\em InterUniversity Centre for Astronomy and Astrophysics,} \\
            {\em Ganeshkhind, Pune 411007 , India.} \\
            }
\renewcommand{\today}{}
\setlength\textwidth{5.75 in}
\setlength\topmargin{-1.cm}
\setlength\textheight{8 in}
\addtolength\evensidemargin{-1.cm}
\addtolength\oddsidemargin{-1.cm}
\font\tenrm=cmr10
\def\baselinestretch{1.4}

\maketitle
\large
\begin{abstract}
The eigen frequencies of radial pulsations of neutron stars are calculated in 
a strong magnetic field. At low densities we use the magnetic BPS 
equation of state(EOS) similar to that obtained by Lai and Shapiro while at 
high
densities the EOS  obtained from the relativistic nuclear mean field theory
is taken and extended to include strong magnetic field. It is found that 
magnetised neutron stars support higher maximum mass where as the effect of magnetic field on radial stability for observed neutron star masses is minimal. 
\end{abstract}
\pagebreak
\begin{section}{introduction}

It is well known that intense magnetic fields(B$\sim$$10^{12-13}$G) exist on 
the surface of many neutron stars.Objects with even higher magnetic fields 
have 
been surmised and detected recently.Recent observational studies and several 
independent arguments link the class of soft $\gamma$-ray repeaters and 
perhaps certain anomalous X-ray pulsars with neutron stars having ultra 
strong magnetic fields, the so called magnetars. Kuoveliotou et al(1998) 
found a soft $\gamma$-ray repeater SGR 1806-20 with 
a period of 
7.47 seconds and a spin down rate of 2.6$\times$$10^{-3}$$syr^{-1}$ from 
which they estimated the pulsar age to be about 1500 years and field strength 
of $\sim$ 8$\times$$10^{14}$G.Since the magnetic field in the core,according 
to some models,could be $10^{3}-10^{5}$ times higher than its value on the 
surface, it is possible that ultra strong magnetic fields of order $10^{18}-10^{19}$G 
or even 
$10^{20}$G exist in the core of certain neutron stars. The virial theorem, it is 
usually argued,gives an upper bound of about $10^{18}$G for the field inside a 
neutron star(Lai and Shapiro 1991). However it has also been 
claimed that this cannot be taken as the last word on the subject, since at 
the super high density inside the star core,general relativistic corrections 
to the virial theorem could increase the upper limit on the maximum allowed 
magnetic fields substantially (Hong 1998). According to Kuoveliotou,a statistical analysis of the population of 
soft $\gamma$-ray repeaters indicates instead of being just isolated examples, as many 
as 10$\%$ of neutron stars could be magnetars. It is therefore of interest to 
study the equation of state of nuclear matter and various properties of 
neutron stars under such magnetic fields. In this paper we study 
the radial oscillations of neutron stars in the presence of super strong 
magnetic fields. Studies of radial oscillations is of interest since Cameron 
suggested (1965) more than three 
decades ago that vibrations of neutron stars could excite motions that can 
have interesting astrophysical applications. X-ray and $\gamma$-ray burst 
phenomenon are clearly explosive in nature. These explosive events probably 
perturb the associated neutron star and the resulting dynamical behaviour may 
eventually be deduced from such observations. Observations of quasi-periodic 
pulses of pulsars have also been associated with oscillations of underlying 
neutron stars (Chandrasekhar 1964 a, b, Cutler et al 1990).  \\
        The EOS is central to the calculation of  most neutron star properties as 
it determines the mass range, the red shift as well as mass-radius 
relationship for these stars. Since 
neutron stars 
span a very wide range of densities, no one EOS is adequate to describe the
properties of neutron stars. In the low density regions from the neutron drip 
density ($\sim$4$\times$$10^{11}$) and upto 
$\rho_n$$\simeq$3.0$\times$$10^{14}$gm/cc the density at which the nuclie 
just begin to dissolve and merge 
together, the nuclear matter EOS is adequately described by the BPS model
(Bayam, Pethick and Sutherland, 1971) which is based on the 
semi-empirical nuclear mass formula. We adopt this BPS EOS and its magnetised 
version  as given by Lai and Shapiro(1991)in this density range. In the high density range above the neutron drip density$\rho_n$, the  
physical properties of matter are still uncertain. Many models for the 
description of nuclear matter at such high densities have been proposed over 
the years. One of the most studied models is the relativistic nuclear mean 
field theory, in which the strong interactions among various particles 
involved are mediated by a scalar field $\sigma$, an isoscalar-vector field 
$\omega$ and an isovector-vector field $\rho$. Along with scalar 
self-intercation 
terms it can reproduce the values of experimentally known quantities revelant 
to nuclear matter, viz. the binding energy per nucleon, the nuclear density at 
saturation, the asymmetry energy,the effective mass and the bulk modulus and 
provides a good description 0f nuclear matter for densities upto a few times 
the saturation density $\rho_c$. In our study, we have used this nuclear mean 
field theory and its modification in the presence of a magnetic field. \\
               In section 2 a brief 
discussion of the EOS is given at zero temperature.In section 3 we present 
the formalism for pulsations of the neutron star models computed here as a 
result of 
integration of the relativistic equations. Section 4 deals with results and 
discussions.
\end{section}

\begin{section}{ The Equation of State (EOS) for Nuclear Matter}
   We shall describe nuclear matter at high densities by the relativistic 
nuclear mean field model, including the $\rho$-contribution, as extended to 
include 
strong interactions. For densities less than the neutron drip density we adopt 
the BPS model in the presence of magnetic field as  developed by 
Lai and Shapiro.

\begin{subsection}{The Nuclear Mean Field EOS at High Densities}
    We consider the charge neutral nuclear matter consisting of neutrons, protons and electrons in $\beta$-equilibrium in the presence of a magnetic field and at zero temperature ($T=0$). The expression for the 
various quantities like, number density, the scalar number density and 
energy-pressure etc in the $[B=0]$ case are well known. In the presence of 
magnetic field these expressions for charged particles are modified in a 
straight forward way, viz
\begin{equation}
\sum_{spin}\int d^{3}p =eB\sum_{\nu=0}^{\nu_{max}}(2-\delta_{\nu{0}})\int dp_{z}
\end{equation}
in the integrals appearing in the expressions for the various quantities 
listed above.
 The total pressure P and the mass energy density ($\rho$) of the system are 
given by
\begin{eqnarray}
P&=&\sum_{i=p,n,e}P_{i}+\frac{1}{2}(\frac{g_{\omega}}{m_{\omega}})^{2}<{\omega_{0}}>^{2}-\frac{1}{2}(\frac{g_{\sigma}}{m_{\sigma}})^{-2}(g_{\sigma}\sigma)^{2}-\frac{1}{3}bm_{n}(g_{\sigma}\sigma)^{3}  \nonumber \\
&-&\frac{1}{4}c(g_{\sigma}\sigma)^{4}+\frac{1}{2}(\frac{g_{\rho}}{m_{\rho}})^{2}<\rho_{0}>^{2}
\end{eqnarray}
\begin{eqnarray}
\rho&=&\sum_{i=p,n,e}\rho_{i}+\frac{1}{2}(\frac{g_{\omega}}{m_{\omega}})^{2}<{\omega_{0}}>^{2}+\frac{1}{2}(\frac{g_{\sigma}}{m_{\sigma}})^{-2}(g_{\sigma}\sigma)^{2}   \nonumber \\
&+&\frac{1}{3}bm_{n}(g_{\sigma}\sigma)^{3}+\frac{1}{4}c(g_{\sigma}\sigma)^{4}+\frac{1}{2}(\frac{g_{\rho}}{m_{\rho}})^{2}<\rho_{0}>^{2}
\end{eqnarray}
In the above
\begin{eqnarray}
P_{n}=\frac{1}{8\pi^{2}}[\frac{1}{3}\mu_{n}^{*}p_{fn}^{*}
(2p_{fn}^{*2}-3m_{n}^{*2})+m_{n}^{*4}
ln(\frac{\mu_{n}^{*}+p_{fn}^{*}}{m_{n}^{*}})]
\end{eqnarray}
\begin{eqnarray}
P_{p}=\frac{eB}{4\pi^{2}}\sum_{\nu=0}^{\nu_{max}}(2-\delta_{\nu0})[\mu_{p}^{*}p_{fp}^{*}-m_{P}^{*2}ln(\frac{\mu_{p}^{*}+p_{fp}^{*}}{m_{p}^{*}})]
\end{eqnarray}
\begin{eqnarray}
P_{e}=\frac{eB}{4\pi^{2}}\sum_{\nu=0}^{\nu_{max}}(2-\delta_{\nu0})[\mu_{e}p_{fe}-m_{e}^{2}ln(\frac{\mu_{e}+p_{fe}}{m_{e}})]
\end{eqnarray}
\begin{eqnarray}
\rho_{n}=\frac{1}{8\pi^{2}}[\mu_{n}^{*}p_{fn}^{*}(2{\mu_{n}}^{*2}-{m_{n}}^{*2})-{m_{n}}^{*4}ln(\frac{\mu_{n}^{*}+p_{fn}^{*}}{m_{n}^{*}})]
\end{eqnarray}
\begin{eqnarray}
{\rho}_{p}=\frac{eB}{4{\pi}^{2}}\sum_{\nu=0}^{\nu_{max}}{(2-{\delta}_{\nu0})}{({\mu_{p}}^{*}{p_{fp}}^{*}+{{m_{p,\nu}}^{*}}^{2}ln{(\frac{{\mu_{p}}^{*}+{p_{fp}}^{*}}{{m_{p}}^{*}})})}
\end{eqnarray}
\begin{eqnarray}
{\rho}_{e}=\frac{eB}{4{\pi}^{2}}\sum_{\nu=0}^{\nu_{max}}{(2-{\delta}_{\nu0})}{(\mu_{e}p_{fe}+{m_{e}}^{2}ln{(\frac{\mu_{e}+p_{fe}}{m_{e}})})}
\end{eqnarray}
\begin{equation}
{{m_{p,{\nu}}}^{*}}^{2}={{m_{p}}^{*}}^{2}+2{\nu}eB
\end{equation}
\begin{equation}
{m_{e,{\nu}}}^{2}={m_{i}}^{2}+2{\nu}eB 
\end{equation}

\begin{equation}
{{p_{fp}}^{*}}^{2}={{{\mu}_{p}}^{*}}^{2}-{{m_{p{\nu}}}^{*}}^{2}
\end{equation}
\begin{equation}
{{p_{fn}}^{*}}^{2}={{{\mu}_{n}}^{*}}^{2}-{{m_{n}}^{*}}^{2}
\end{equation}
\begin{equation}
{p_{fe}}^{2}={{\mu}_{e}}^{2}-{m_{e}}^{2}-2{\nu}eB
\end{equation}
\begin{equation}
m_{p}^{*}-m_{p}={m_{n}}^{*}-m_{n}=-{(\frac{g_{\sigma}}{m_{\sigma}})}^{2}{n_{s}}
\end{equation}
\begin{equation}
n_{s}=n_{p}^{s}+{n_{n}}^{s}
\end{equation}
\begin{equation}
{n_{n}}^{s}=\frac{m_{n}^{*}}{2{\pi}^{2}}{({{\mu}_{n}}^{*}p_{fn}-{{m_{n}}^{*}}^{2}ln{(\frac{{{\mu}_{n}}^{*}+{p_{fn}}^{*}}{{m_{n}}^{*}})})}
\end{equation}
\begin{equation}
n_{p}^{s}=\frac{eB{m_{p}^{*}}}{2{\pi}^{2}}\sum_{\nu=0}^{\nu_{max}}{(2-{\delta}_{{\nu},0})}ln{(\frac{{{\mu}_{p}}^{*}+{p_{fp}}^{*}}{{m_{p,{\nu}}}^{*}})}
\end{equation}
\begin{equation}
n_{B}=n_{p}+n_{n}
\end{equation}
\begin{equation}
n_{n}=\frac{{{p_{fn}}^{*}}^{3}}{3{{\pi}^{2}}}
\end{equation}
\begin{equation}
n_{p}=\frac{eB}{2{\pi}^{2}}\sum_{\nu=0}^{\nu_{max}}{(2-{\delta}_{{\nu},0})}{p_{fp}}^{*}
\end{equation}
\begin{equation}
n_{e}=\frac{eB}{{2{\pi}}^{2}}\sum_{\nu=0}^{\nu_{max}}{(2-{\delta}_{\nu,0})}{p_{fe}}
\end{equation}
\begin{equation}
n_{p}=n_{e}
\end{equation}
\begin{equation}
<\rho_{0}>=\frac{1}{2}(n_{P}-n_{n})
\end{equation}
\begin{equation}
<\omega_{0}>=n_{B}
\end{equation}
\begin{equation}
\mu_{i}^{*}=\mu_{i}-(\frac{g_{\omega}}{m_{\omega}})^{2}<\omega_{0}>-I_{3i}(\frac{g_{\rho}}{m_{\rho}})^{2}<\rho_{0}>
\end{equation}  \\
for i=n,p \\
\begin{equation}
{\mu}_{n}={\mu}_{p}+{\mu}_{e}
\end{equation}
For a given value of $\mu_e$ and B we can find  self consistently the values of
$m_{p}^{*}$, $\mu_{p}^{*}$, $\mu_{n}^{*}$ and so the pressure and mass energy 
can be 
computed. This helps us to compute the equation of state in the mean field 
approximation as a function of $\mu_{e}$ or $n_{B}$.
\end{subsection}

\begin{subsection}{The Magnetic BPS model}
   The total pressure of the hadron matter below the neutron drip is given 
by
\begin{equation}
P=P_{e}{(n_{e})}+P_{L}=P_{e}{(n_{e})}+\frac{1}{3}{\varepsilon}_{L}{(Z,n_{e})}
\end{equation}
where ${\varepsilon}_{L}$ is the bcc coulomb lattice energy and $P_{e}$ the 
pressure of the electrons in the presence of the magnetic field is given by
\begin{eqnarray}
P_{e}&=&\frac{eB}{4\pi^{2}}\sum_{\nu=0}^{{\nu}_{max}}(2-{\delta_{\nu0}})[\mu_{e}\sqrt{\mu_{e}^{2}-m_{e}^{2}-2\nu eB}  \nonumber \\
&-&(m_{e}^{2}+2\nu eB)ln(\frac{\mu_{e}+\sqrt{\mu_{e}^{2}-m_{e}^{2}-2\nu eB}}{\sqrt{m_{e}^{2}+2\nu eB}})]
\end{eqnarray}
\begin{equation}
P_{L}=-1.444{Z}^{\frac{2}{3}}e^{2}e^{2}{n_{e}}^{\frac{4}{3}}
\end{equation}
\begin{equation}
n_{e}=\frac{eB}{4{\pi}^{2}}\sum_{\nu=o}^{{\nu}_{max}}{(2-{{\delta}_{{\nu}0}})}{[{{\mu}_{e}}^{2}-{m_{e}^{2}}-2{\nu}eB]}^{\frac{1}{2}}
\end{equation}
Consider matter condensing into a perfect crystal lattice with a single 
nuclear species(A,Z) at the lattice sites.The energy density is
\begin{equation}
\varepsilon=\frac{n_{e}}{Z}W_{N}{(A,Z)}+\varepsilon_{e}^{'}{(n_{e})}+{\varepsilon}_{L}{(Z,n_{e})}
\end{equation} 
where $W_{N}$ is the mass-energy of the nucleus(including the rest mass of 
nucleons and Z electrons) $\varepsilon_{e}^{'}$ is the free electron energy 
including the rest mass of electrons.Following Lai and Shapiro we use for 
$W_{N}$, the experimental values for laboratory nuclei as tabulated by 
Wapastra and Bos(1976,1977). The elements taken in this paper are given below in the 
table along with their mass energy $W_{N}(A,Z)$. At a given pressure P,the 
equilibrium values of A and Z are determined by minimising the Gibbs free 
energy per nucleon,
\begin{equation}
g=\frac{{\varepsilon}+P}{n}=\frac{W_{N}{(A,Z)}}{A}+\frac{Z}{A}{({\mu}_{e}-m_{e}c^{2})}+\frac{4{\varepsilon}_{L}}{3A{n_{e}}}
\end{equation}
The neutron drip point is determined by the condition
\begin{equation}
g_{min}=m_{n}c^{2}
\end{equation}
Knowing A and Z the energy can be determined from Eq.(32)
\end{subsection}
\end{section}

\begin{section}{Radial Pulsations of a Non-Rotating Neutron Star}

The equations governing infinitesimal radial pulsations of a non-rotating 
star in general relativity were given by Chandrasekhar(1964). The structure of 
the star in hydrostatic equilibrium is described by the Tolman-Openheimer-
Volkoff equations
\begin{equation}
\frac{dp}{dr}=\frac{-G{(p+{\rho}c^{2})}{(m+\frac{4{\pi}r^{3}p}{c^{2}})}}{c^{2}r^{2}{(1-\frac{2Gm}{c^{2}r})}}
\end{equation}
\begin{equation}
\frac{dm}{dr}=4{\pi}{r^{2}}{\rho}
\end{equation}
\begin{equation}
\frac{d{\nu}}{dr}=\frac{2Gm{(1+\frac{4{\pi}r^{3}p}{mc^{2}})}}{c^{2}r{(1-\frac{2Gm}{c^{2}r})}}
\end{equation}
Given an equation of state $p(\rho)$,equations (35)-(37) can be integrated 
numerically for a given central density to obtain the radius R and 
gravitational mass $M=M(R)$ of the star. The metric used is given by
\begin{equation}
d{s^{2}}=-{e^{2}}^{\nu}{c^{2}}{d{t^{2}}}+{e^{2}}^{\lambda}{d{r^{2}}}+{r^{2}}{(d{\theta}^{2}+{sin^{2}}{\theta}d{\phi}^{2})}
\end{equation}
If ${\Delta}r$ is the radial displacement
\begin{equation}
\xi=\frac{{\Delta}r}{r}
\end{equation}
\begin{equation}
\zeta=r^2{e^{-}}^{\nu}{\xi}
\end{equation}
and the time dependence of the harmonic oscillations is written as 
$exp(i{\sigma}t)$,one gets the equation governing radial adiabatic 
oscillations(Chandrasekhar 1964,Datta etal 1998,Anand etal 2000)
\begin{equation}
F\frac{d^{2}{\zeta}}{d{r^{2}}}+G\frac{d{\zeta}}{dr}+H{\zeta}={\sigma}^{2}{\zeta}
\end{equation}
where
\begin{equation}
F=-\frac{e^{2{\nu}-2{\lambda}}{({\Gamma}p)}}{p+{\rho}c^{2}}
\end{equation}
\begin{equation}
G=-\frac{e^{2{\nu}-2{\lambda}}}{p+{\rho}c^{2}}{[{({\Gamma}p)}{({\lambda}+3{\nu})}+\frac{d}{dr}{({\Gamma}p)}-\frac{2}{r}{({\Gamma}p)}]}
\end{equation}
\begin{equation}
H=\frac{e^{2{\nu}-2{\lambda}}}{p+{\rho}c^{2}}{[\frac{4}{r}\frac{dp}{dr}+\frac{8{\pi}G}{c^{4}}{e^{2}}^{\lambda}p{(p+{\rho}c^{2})}-\frac{1}{p+{\rho}c^{2}}{(\frac{dp}{dr})}^{2}]}
\end{equation}
\begin{equation}
\lambda=-ln{[1-\frac{2Gm}{r{c^{2}}}]}^{\frac{1}{2}}
\end{equation}
In the above equations $\Gamma$ is the adiabatic index given by \\
\begin{equation}
{\Gamma}=\frac{p+{\rho}c^{2}}{c^{2}p}\frac{dp}{d\rho}
\end{equation}
The boundary conditions to solve equation (41)  are
\begin{eqnarray} 
\zeta (r=0)=0  \nonumber \\
\delta p(r=R)=0 
\end{eqnarray}
The expression for ${\delta}p$ as given by Chandrasekhar(1964) is
\begin{equation}
{\delta}p{(r)}=-\frac{dp}{dr}\frac{e^{\nu}{\zeta}}{r^{2}}-\frac{{\Gamma}p{e^{\nu}}}{r^{2}}\frac{d{\zeta}}{dr}
\end{equation}
All these equations are totally model independent and are infact the same whether we 
are considering neutron stars, quark stars or any other dense stellar object. The nature of the object being considered and the particular model affects 
the structure of the star and the frequency of radial pulsations only through 
the EOS. Notice that in Chandrasekhar (1964) and Datta etal (1998) the pulsation equations were written in terms of  $\xi$ instead of $\zeta$. Equation (42) along with the boundary conditions represent a Sturm-Liouville eigenvalue problem for ${\sigma}^{2}$. From the theory of such equations we have the well known results:
(i)  Eigenvalues ${\sigma}^{2}$ are all real, and
(ii) They form an infinite discreet sequence\\
     $~~~~~~~~~\sigma_{0}^{2} < \sigma_{1}^{2} < \sigma_{2}^{2}$ ....\\	
An important consequence of (ii) is that if the fundamental radial mode of a star is stable $({\sigma_{0}}^{2} > 0)$, then all the radial modes are stable.
\end{section}
\begin{section}{Results and Discussions}
To study the structure and radial oscillations of neutron stars in the presence of a strong magnetic field we have employed the BPS model with its generalization in a magnetic field given by Lai and Shapiro(1991) below the neutron drip and the RMF theory above it. We have used the values of various couplings fixed by
Ellis et al(1991) which provide the known values of various nuclear matter parameters :
\begin{eqnarray}
(\frac{g_{\sigma}}{m_{\sigma}})&=&0.01525 Mev^{-1}  \nonumber  \\
(\frac{g_{\omega}}{m_{\omega}})&=&0.011 Mev^{-1}     \nonumber  \\
(\frac{g_{\rho}}{m_{\rho}})&=&0.011 Mev^{-1}          \nonumber \\
b&=&0.003748                                         \nonumber  \\
c&=&0.01328                                          \nonumber  \\
\end{eqnarray}
As explained in section 2.1 and 2.2, for the RMF theory the equations are solved in a self-consistent manner for the effective masses and chemical potentials ,and hence the EOS. Below the neutron drip, the EOS is obtained by the minimization of Gibb's free energy as functions of A and Z. For this purpose we have employed 14 nuclei listed in the table. The problem is solved seprately for $B=0$ and $B\neq0$. For each B this gives the EOS in the form $P(n_{B})$ and $\rho(n_{B})$. The structure of the neutron star is then obtained from the integration of the oppenheimer-Volkoff equations. This then gives the profile of m,p and $\nu$ as a function of r for each star. One more quantity that is required is $\Gamma$ which is then calculated directly from the EOS for all densities by using a quadratic difference formula for the derivative $\frac{dp}{d(\rho c^{2})}$.  \\
             Along with the M-R relationship, one also obtains the gravitational red shift
\begin{equation}
Z=[1-\frac{2Gm}{c^{2}r}]^{-\frac{1}{2}}-1
\end{equation}  \\
which can in principle be observed experimentally.\\
            The procedure to obtain the eigen frequencies is simple. We guess a value for $\sigma$ and integrate the equation outword from the centre upto the surface. The guessed value of $\sigma$ is varied till the boundary condition \\
              $~~~~~~~~~~~~~~~~~~~~~\delta p(r)=0$ at $r=R$ \\
is satisfied. 
\end{section}

In figure 1. we plot mass in solar mass unit vs radius in Kms. for magnetic fields 0, 1$\times$$10^{3}$, 5$\times$$10^{4}$, 1$\times$$10^{5}$ $MeV^{2}$ $(1MeV^{2}=1.69\times10^{14}$G) represented by the curves A, B, C and D respectively. It is worthwile to note that the magnetised neutron stars support higher masses. For very high magnetic fields the stars become relatively more compact. In figure 2. we present mass vs central energy density and in figure 3. we have plotted gravitational red shift vs mass. In figure 4. a plot of time period of fundamental mode vs gravitational red shift is given for the same magnetic fields as in figure 1.
It is interesting to note that for the observed neutron star mass(1.4 $M_{\odot}$), the magnetic field has pratically no influence on radial stability. Similar trend is seen for the first excited mode as shown in figure 5.
\pagebreak

\begin{figure}[ht]
\vskip 15truecm
\includegraphics{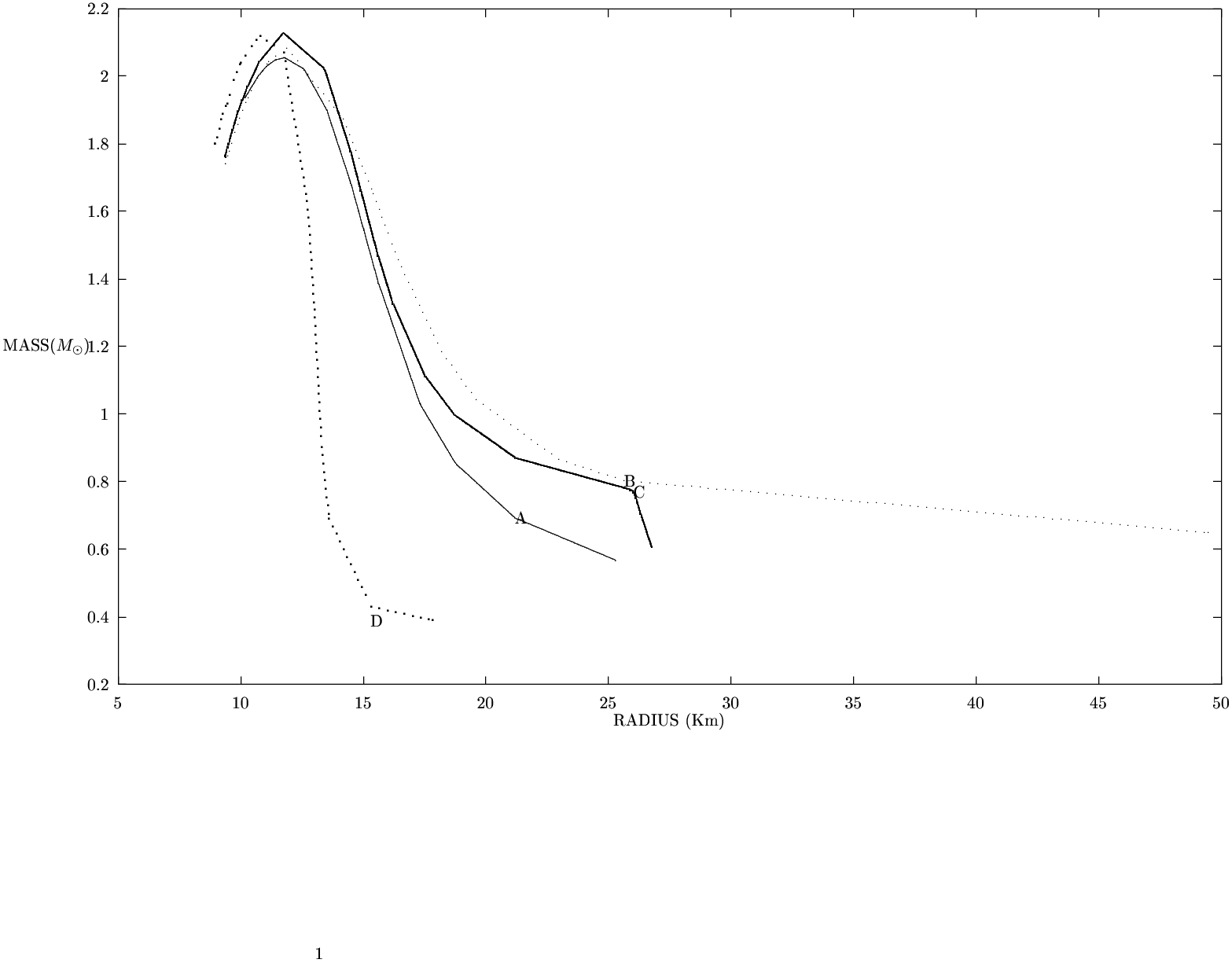}
\caption{Plot of mass in solar mass unit vs radius in Kms. for magnetic fields 0, 1$\times$$10^{4}$, 5$\times$$10^{4}$, 1$\times$$10^{5}$ $MeV^{2}$ represented by the curves A, B, C and D respectively.}
\end{figure}

\begin{figure}[ht]
\vskip 15truecm
\includegraphics{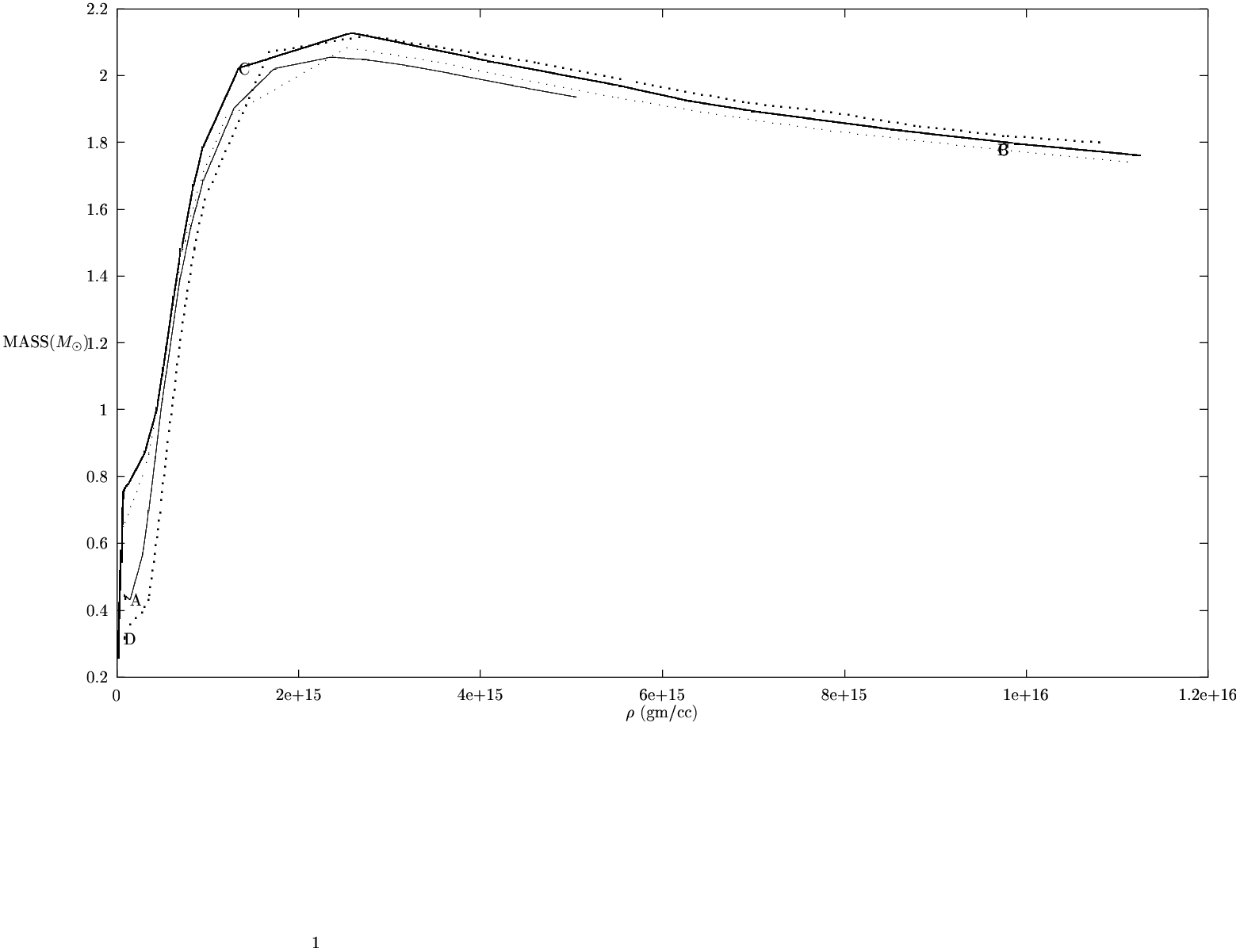}
\caption{Plot of mass in solar mass unit vs energy density for magnetic fields 0, 1$\times$$10^{4}$, 5$\times$$10^{4}$, 1$\times$$10^{5}$ $MeV^{2}$ represented by the curves A, B, C and D respectively.}
\end{figure}

\begin{figure}[ht]
\vskip 15truecm
\includegraphics{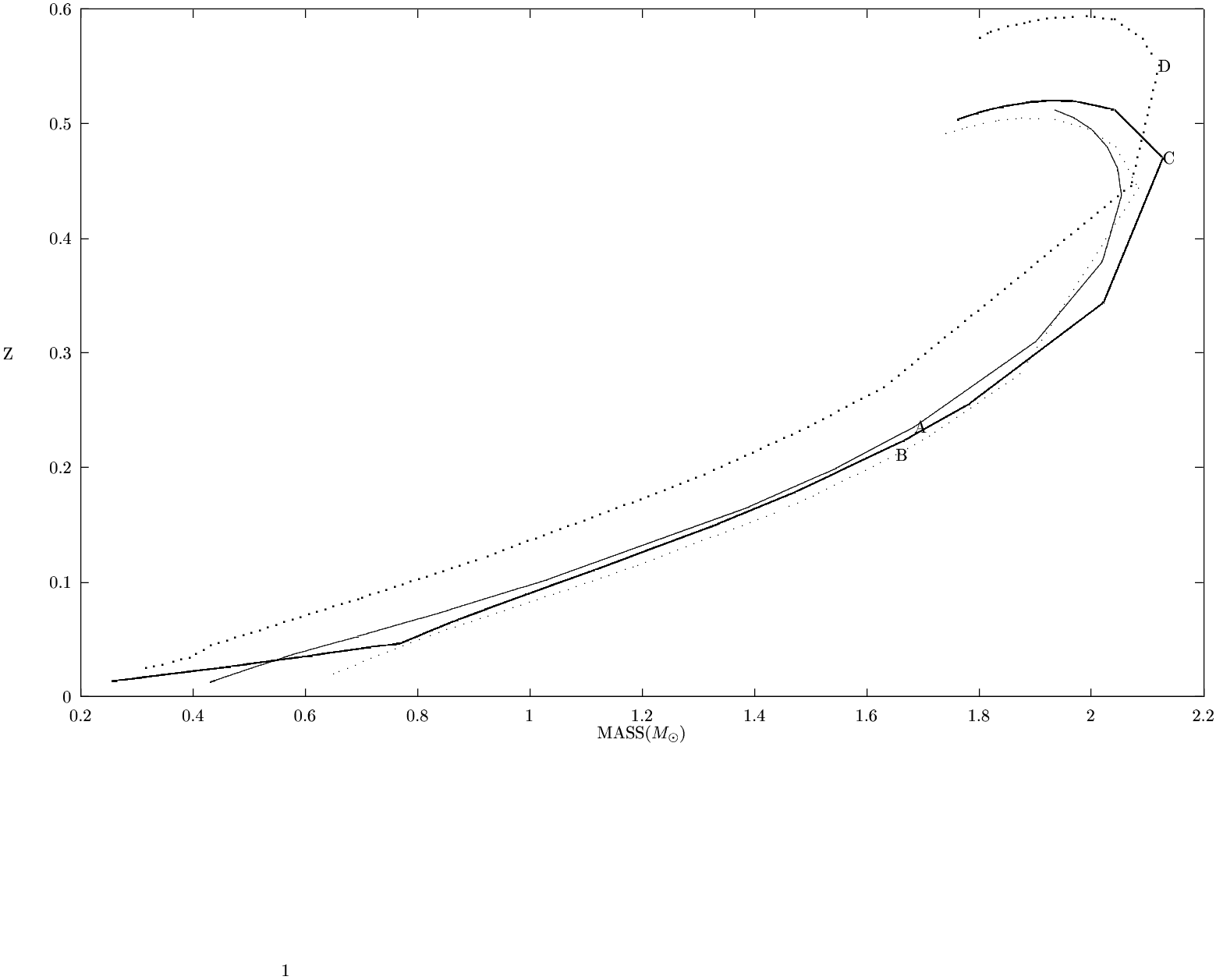}
\caption{Plot of gravitational red shift(Z) vs mass in solar  mass unit for magnetic fields 0, 1$\times$$10^{4}$, 5$\times$$10^{4}$, 1$\times$$10^{5}$ $MeV^{2}$ represented by the curves A, B, C and D respectively.} 
\end{figure}

\begin{figure}[ht]
\vskip 15truecm
\includegraphics{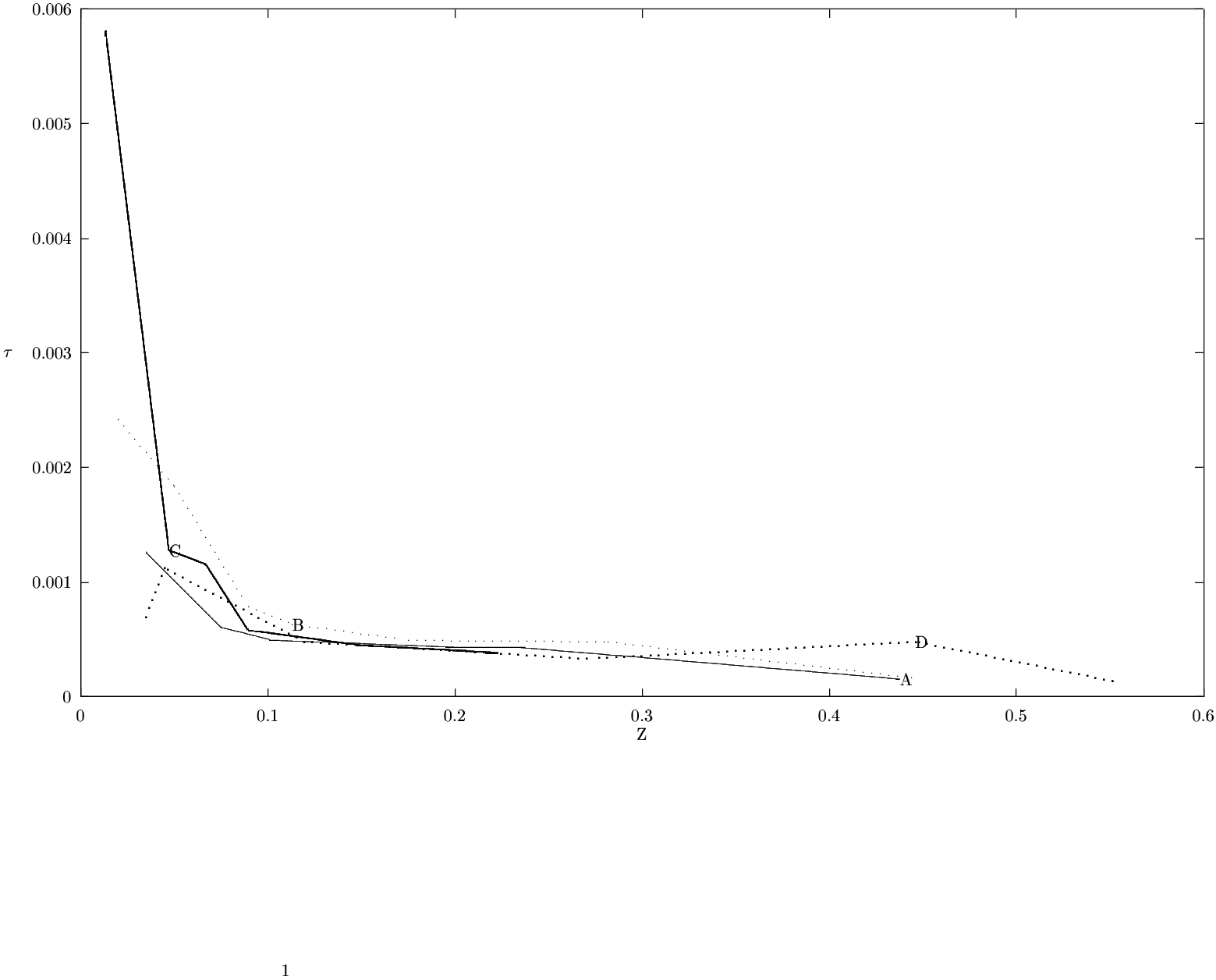}
\caption{Plot of time period $\tau$ for fundamental mode vs gravitational red shift(Z) for magnetic fields 0, 1$\times$$10^{4}$, 5$\times$$10^{4}$, 1$\times$$10^{5}$ $MeV^{2}$ represented by the curves A, B, C and D respectively.} 
\end{figure} 

\begin{figure}[ht]
\vskip 15truecm
\includegraphics{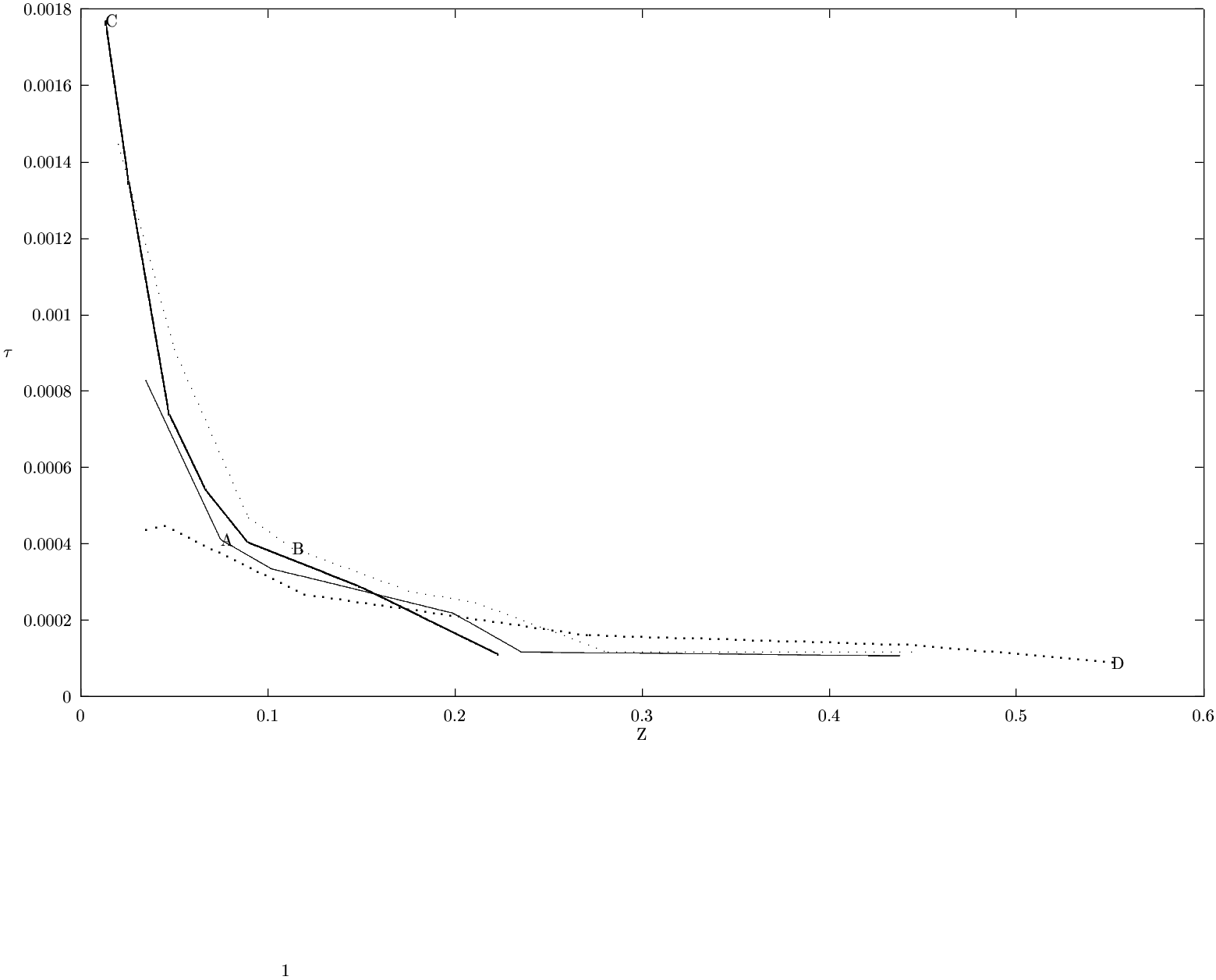}
\caption{Plot of time period $\tau$ for n=1 mode vs gravitational red shift(Z) for magnetic fields 0, 1$\times$$10^{4}$, 5$\times$$10^{4}$, 1$\times$$10^{5}$ $MeV^{2}$ represented by the curves A, B, C and D respectively.}
\end{figure} 
\pagebreak
\begin{center}
TABLE: BPS Equilibrium Nuclei Below Neutron Drip \\
\begin{tabular}{|c|c|} \hline 
 Element         & BPS MASS-ENERGY \\
                 & (in units of $10^{4}$ Mev)\\  \hline\hline
$ Fe_{26}^{56} $ & 5.2103                   \\ 
$ Ni_{28}^{62} $ & 5.7686                    \\
$ Ni_{28}^{64} $ & 5.9549                     \\
$ Ni_{28}^{66} $ & 6.1413                      \\
$ Kr_{36}^{86} $ & 8.0025                       \\
$ Se_{34}^{84} $ & 7.8170                        \\
$ Ge_{32}^{82} $ & 7.6316                      \\
$ Zn_{30}^{80} $ & 7.4466                      \\
$ Ni_{28}^{78} $ & 7.2621                      \\
$ Ru_{44}^{126}$ & 11.7337                     \\
$ Mo_{42}^{124}$ & 11.5495                     \\
$ Zr_{40}^{122}$ & 11.3655                     \\
$ Sr_{38}^{120}$ & 11.1818                     \\
$ Sr_{38}^{122}$ & 11.3655                     \\
$ Kr_{36}^{118}$ & 10.9985                     \\ \hline

\end{tabular}
\end{center}
\end{document}